\documentstyle[epsf]{mn}
\input psfig.sty

\title{X-ray Spectra of the RIXOS source sample}
\author[J. P. D. Mittaz, et al.]
       {J. P. D. Mittaz$^1$, F.J. Carrera$^{1,2}$, E. Romero-Colmenero$^1$,
        K. O. Mason$^1$, \and  G. Hasinger$^3$, 
	R. McMahon$^4$, H. Andernach$^{5}$, R. Bower$^{6}$, J. 
	Burgos-Martin$^7$, 
	\and J. I. Gonz\'alez--Serrano$^2$, D. Wonnacott$^1$ \\
	$^1$Mullard Space Science Laboratory, University College London, 
	Holmbury St. Mary, Dorking, Surrey, RH5 6NT\\
        $^2$Instituto de F\'\i sica de Cantabria,
          Consejo Superior de Investigaciones
          Cient\'\i ficas - Universidad de Cantabria, Santander, Spain\\
	$^3$Astrophysikalisches Institut Potsdam, An der Sternwarte 16 , 
	Potsdam, Germany \\
        $^4$Institute of Astronomy, Madingley Road, Cambridge, U.K. \\
	$^5$Depto. de Astronom\'\i a, IFUG, Guanajuato, Mexico\\
        $^6$Department of Physics, University of Durham, South Road, Durham\\
	$^7$ Instituto de Astrofisica de Canarias, La Laguna, Tenerife, Spain\\
	}

\date{Accepted ;
      Received ;
      in original form 1997 March 27}

\begin{document}

\maketitle

\begin{abstract}

We present results of an extensive study of the X-ray spectral properties of
sources detected in the RIXOS survey, which is a large,  nearly complete
sample of objects detected serendipitously in ROSAT PSPC fields down to a flux
limit of $3 \times 10^{-14}$ ergs cm$^{-2}$ s$^{-1}$ (0.5 -- 2 keV).  We show
that for X-ray surveys containing sources with low count rate,  such as RIXOS,
spectral slopes estimated using simple hardness ratios in the ROSAT band can be
biased.  Instead  we analyse three-colour X-ray data using statistical
techniques appropriate to the Poisson regime which removes the effects of this
bias. We also show that the use of three-colour data enables some 
discrimination between thermal and non-thermal spectra.  We have then
applied this technique to the RIXOS survey to study the spectral properties of
the sample.

For the AGN we find an average energy index of $1.05 \pm 0.05$ with no evidence
for spectral evolution with redshift.  Individual AGN are shown to have a
range of properties including soft X-ray excesses and intrinsic absorption. 
Narrow Emission Line Galaxies (NELGs) also seem to fit to a power-law spectrum,
which may indicate a non-thermal origin for their X-ray emission.   We infer
that most of the clusters in the sample have a bremsstrahlung temperature $> 3$
keV, although some show evidence for a cooling flow.  The stars deviate
strongly from a power-law model but fit to a thermal model.  Finally, we have
analysed the whole RIXOS sample (extending the flux cutoff to the
sensitivity threshold of each individual observation)  containing 1762 sources
to study the relationship between spectral slope and flux. We find that the
mean  spectral slope of the sources hardens at lower fluxes in agreement with
results from other samples. However,  a study of the individual sources
demonstrates that the majority  have relatively soft spectra even at faint flux
levels, and the hardening of the mean  is caused by the appearance of  a
population of very hard sources at the lowest fluxes. This has implications for
the nature of the soft X-ray background.

\end{abstract}
\begin{keywords}
surveys - galaxies:active - quasars:general - X-rays:galaxies - X-rays:stars
\end{keywords}

\section{Introduction}

X-ray surveys have proven to be powerful tools in extending our knowledge of a
range of object types, from highly luminous AGN to active stars.  Surveys of
`serendipitous' detections in the fields of view of imaging X-ray instruments,
examples of which are the Einstein Medium Sensitivity Survey (EMSS, Gioia et
al. 1990) and the EXOSAT High Galactic Latitude Survey (HGLS, Giommi et al.
1991), have provided large samples with which to make detailed statistical
studies with relatively well-understood selection biases.  With the advent of
ROSAT ever more extensive and sensitive surveys are becoming available, ranging
from the ROSAT all sky survey which sampled relatively bright source
populations, to deep pencil-beam surveys such as those of Hasinger et al.
(1993) and Branduardi-Raymont et al. (1994).  Other samples have
concentrated on serendipitous sources discovered in ROSAT pointed data
(e.g. Boyle et al. 1994, Carballo et al. 1996, Boyle et al. 1995).

The spectral properties of such samples can be a crucial element in
understanding the nature of the X-ray emission.  However, much of the work to
date has considered only the broad-band fluxes of survey sources. The subject
of this paper is the RIXOS survey of ROSAT field sources which covers a total
of 20~deg$^2$ of sky and has a high level of optical identification
completeness ($\sim$94\% over a 15~deg$^2$ sub-area) down to a flux level $3
\times 10^{-14}$ ergs cm$^{-2}$ s$^{-1}$. This flux cutoff is set so as to
bridge the gap in sensitivity and sky coverage between the ROSAT all-sky survey
and the deepest pencil-beam ROSAT surveys. As the flux cutoff of RIXOS is set
at a level which is much higher than the sensitivity threshold of the ROSAT
observations used, sufficient numbers of X-ray photons have been detected from
all RIXOS sources to provide some information about their overall spectral
distribution. This paper examines the X-ray spectral properties of the RIXOS
sample.  Other aspects of the RIXOS survey are covered in Mason et al. (1998),
Page et al. (1996), Puchnarewicz et al. (1996,1997), Carrera et al. (1998) and
Romero-Colmenero et al. (1998).  The paper discussing cluster evolution by
Castander et al. (1995) is based on a subset of the RIXOS complete sample.

\section{The RIXOS Sample}\label{sec-sample}

The X-ray data are taken from the RIXOS sample of objects (Mason et al. 1998)
and have been constructed from serendipitous sources discovered in 80 pointed
ROSAT PSPC fields.  The fields were chosen to have nominal exposure times
greater than 8 ksec and to be above a Galactic latitude of 28 degrees.  This
limit enables us to sample sources at faint fluxes without the problem of
identifying them in crowded fields.  From each field we have excluded the
target of the observation and only consider sources at less than 17 arc minutes
off-axis.  Such sources have the best positional certainty and are not masked
by the detector window support structure.  Survey sources are selected in the
0.4 - 2 keV band; the poorer point spread function and increased background
due to diffuse Galactic X-ray emission make the detection of X-ray sources more
difficult at softer energies.

Full details of the optical imaging and spectroscopy and
identification process are given in Mason et al. (1998). Over 82
fields (or 20.3 square degrees) our sample is completely identified down
to a flux limit of $8.4 \times 10^{-14}$ ergs cm$^{-2}$ s$^{-1}$ and
over 61 fields (or 14.9 square degrees) we have complete
identifications down to our target flux limit of $3 \times 10^{-14}$
ergs cm$^{-2}$ s$^{-1}$.  This flux limit is well above the detection
limit for all our fields, and for many sources gives a reasonable
number of observed counts. Table 2 lists all the sources in the RIXOS
fields above a flux limit of $3 \times 10^{-14}$ ergs cm$^{-2}$
s$^{-1}$ (0.5 - 2 keV), giving field ID and source ID (for details see
Mason et al. 1998) together with the galactic column ($N_H$), date of
observation and exposure time (column 5). This is the sample with
which we are primarily concerned here, and it will be referred to here
as RIXOS.

 In total the RIXOS sample contains 404 sources, of which 347 have
been identified.  The identification of the sources has been based
largely on the optical spectra, and we have split them into six
categories. These are Active Galactic Nuclei (AGN), Narrow Emission
Line galaxies (NELGs, which may include Seyfert 2 galaxies, LINERs,
and HII region galaxies), isolated galaxies, clusters of galaxies,
active stars and dMe stars.  Of the 347 identified sources, 16 are so
close together that no separate spectra could be extracted for them,
their spectra are included in Table 2 as `MERG'. Five more sources
(one of them unidentified) were in fields 115 and 116 (Mason et
al. 1998) for which no public archival X-ray data were available at
the time of writing. This leaves us with 327 identified sources with
available X-ray data, of which 205 have been classified as AGN, 18 as
NELGs, 6 as isolated galaxies, 30 as clusters, 46 as stars and 22 as
dMe stars.  In addition, we have also fitted the spectra of 56
unidentified sources (included in Table 2 as `UNKN').  In total, the
RIXOS sample forms the largest serendipitous survey constructed from
ROSAT PSPC pointings to date, with a larger sky coverage than
comparable samples such as the Cambridge-Cambridge ROSAT Serendipity
Survey.

\section{Data reduction}

From the RIXOS sample we have taken all those sources which have a firm
optical identification and have extracted three-colour X-ray data.  After the
recommendation of Snowden et al.  (1993) we have used bands S1 (channels 8-41),
H1 (channels 52-90) and H2 (channels 91-201).  For each field we have
constructed an image in each of the three-colours and have ensured the optimal
signal-to-noise by excluding high background times and those times when the
attitude solution was bad.  In general, this excluded between 5 and 20\% of the
data.  We then extracted the source counts for all the known sources in the
field (including those with no identification) using an extraction circle of 54
arc seconds, which includes 90\% of the ROSAT PSF and maximises the signal to
noise for weak sources. In those cases where there was a contaminating source
nearby, the extraction circles were reduced in size until there was no overlap.
The sizes of the extraction circles for each source are listed in
table 2 (column 6) and the fraction of the PSF included for each
source is taken into account during the spectral fitting process (section 5.1).

As we are studying very faint sources, we have been careful to obtain an
accurate estimate of the background.  After masking out all the sources from an
image, it was flattened using the exposure map supplied as part of the standard
SASS processing. As the exposure map corrects for vignetting and other
instrumental effects, we can obtain an accurate estimate of the image
background corrected for systematic instrumental effects by summing over a
large number of pixels.  For RIXOS we summed the data between 5.2 and 10.8
arc-minutes off-axis thereby excluding the residual effects of any bright
central source.  We can then estimate the background at any given source
position from the mean background using data from the exposure map at the
required position.  This method yields a very accurate background estimate
based on a very large number of pixels compared to the number of pixels in the
source extraction circle and to a very good degree of approximation we can then
assume that this background estimate has a negligible error.  Table 2 lists the
the extracted counts for the RIXOS sample (columns 8-10) and the
background estimates in each of the three bands (columns 11-13).

\section{The Colour-Colour diagram}\label{colour-colour-sec}

\begin{figure*}
\psfig{figure=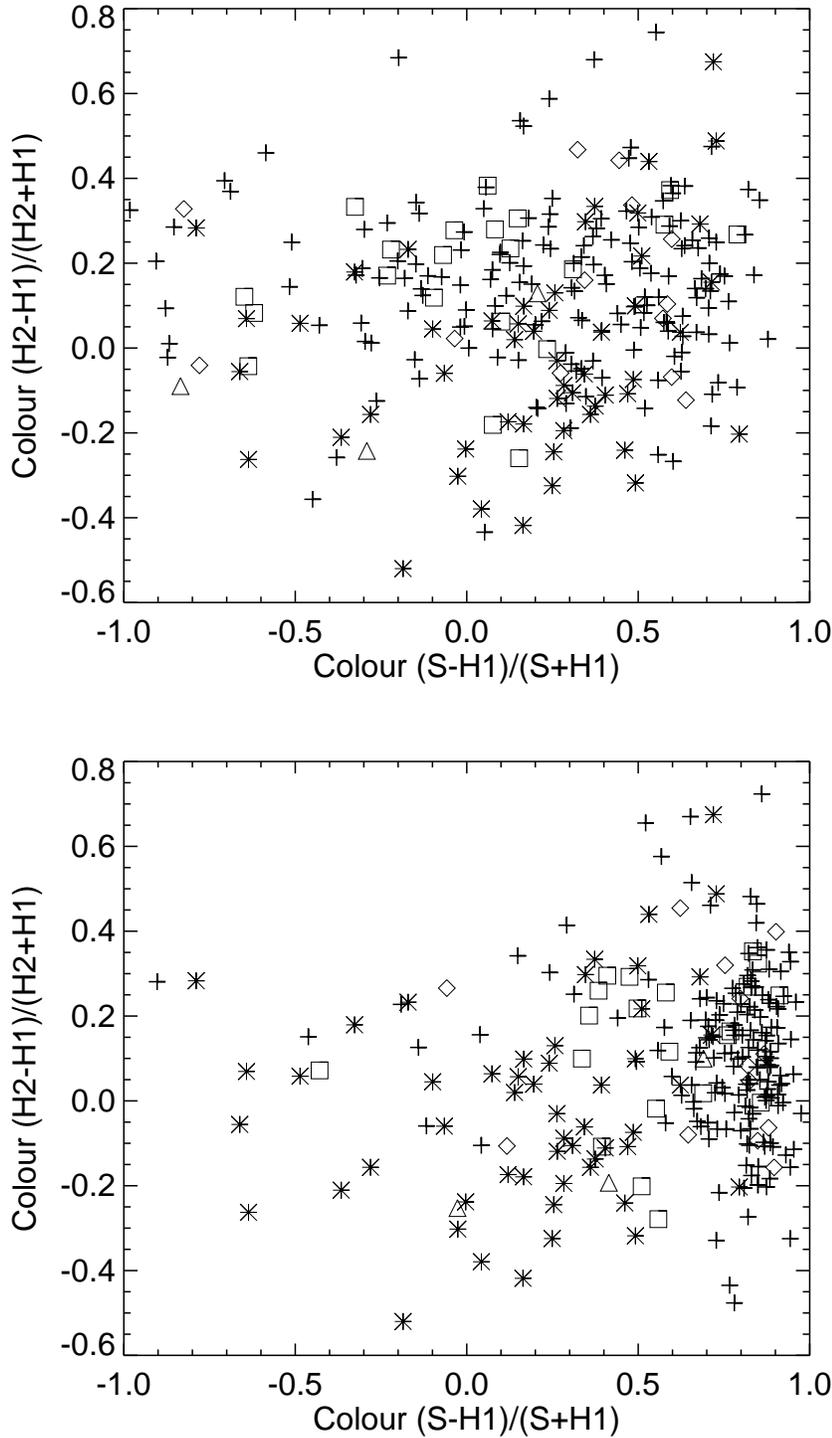,height=20truecm,width=12truecm}
\caption{The Colour-Colour diagrams for the RIXOS sample of objects.  The top
panel shows the data uncorrected for the effect of galactic $N_H$, while the
bottom panel is corrected for galactic $N_H$.  Different classes of objects
have different symbols ($+$ AGN, $\diamond$ ELG, $\bigtriangleup$ Galaxies,
$\Box$ clusters and $\star$ stars (including dMe stars)).}
\label{fig1}
\end{figure*}

As a first step in studying the spectral properties of the sources, we 
have constructed a colour-colour diagram including all identified
sources in the RIXOS sample (Figure~\ref{fig1}).  Our
normalised colours are defined as

\begin{equation}
C1 = (S-H1)/(S+H1)
\end{equation}

\noindent and 

\begin{equation}
C2 = (H2-H1)/(H2+H1).
\end{equation}

The first plot in figure~\ref{fig1} shows uncorrected colours, and the second
shows colours corrected for the effect of galactic absorbing column where
applicable. As the correction for the galactic column is model dependent, we
have used a power-law fitted to the three-colours (see section~\ref{sec-fits}),
and have only applied the correction to extra-galactic sources.
Figure~\ref{fig1} shows a number of features.  On average the AGN tend to be
softer than most of the other sources when corrected for galactic absorption,
though it is clear that not all the AGN are soft and some AGN occupy portions
of the diagram appropriate to hard sources (see also Figure~\ref{fig-hist}).
Five out of 205 AGN have $C1 < 0$, implying that they are very hard or
intrinsically absorbed. A further 15 sources do not appear on this diagram at
all since they were not detected in the soft band, and of these six are
identified with AGN.  This implies that $\sim 5$\% of AGN are very hard and are
candidates for intrinsic absorption.

As a first step in categorising the spectral characteristics of our sample, we
have quantified the differences between the different types of X-ray sources in
the colour-colour plot by using a two-dimensional Kolmogorov-Smirnov test. The
method used is taken from Press et al. (1992) and the results are shown
in Table~\ref{table_ks} for both the un-corrected and corrected colour-colour
data.  The probabilities quoted are of the two samples being drawn from the
same parent distribution.  However, many of the sources are faint and therefore
have large uncertainties which are not taken into account by a standard KS
test.  We have quantified the possible effect of the measurement uncertainties
on the K-S probabilities.  To do this we have simulated 100 samples with the
same flux distribution as our real sample but assuming that all sources have
power-law slopes distributed in a similar way to the AGN.  We have used a mean
slope of $\alpha = 1$ and a dispersion of 0.55.  The numbers in brackets in
table~\ref{table_ks} are the fraction of the time that a K-S probability was
obtained that was smaller than the one seen in the original dataset.  This
therefore gives an indication of the likelihood of obtaining a probability as
small as that seen or better by chance alone given our assumption concerning
the distribution of AGN slopes.

The 2-d Kolmogorov-Smirnov test emphasises the fact that on average the AGN and
objects classified as Narrow Emission Line Galaxies (NELGs) lie in a region of
the colour-colour plot distinct from the clusters and stars.  The similarity
between the NELGs and the AGN as well as the disparity between the NELGs and
the clusters/stars is intriguing and may hint at NELGs containing AGN like
activity.  If the emission were to arise solely from thermal emission from hot
gas, the NELGs may be expected to lie further to the left in the colour-colour
diagram, closer to the clusters.

`

The other sources lie to the left of the AGN in the colour-colour diagram.
That the stars and clusters are distinct from the AGN is not surprising, given
the different physical mechanism known to underly their X-ray emission.  From
Figure~\ref{fig1} the stars constitute the hardest sources with the clusters
lying midway between the stars and AGN.  However, within the stars from
Table~\ref{table_ks} there is a further difference which would seem to indicate
that the dMe stars are softer than other active stars.  Given
the multi-temperature nature of the emission from stars, simple three-colour
data cannot give more than an indication of a difference in the X-ray spectra
between these two classes of objects.  

\begin{table}
\caption{Two dimensional Kolmogorov-Smirnov probabilities for the
different classes of objects in the sample based on the both the
un-corrected and corrected colour-colour diagram.  The numbers in brackets
are the associated probabilities based on simulated datasets.}\label{table_ks}
\begin{tabular}{|l|l|c|c|} \hline
Source 1 type & Source 2 type & 2-d KS & 2-d KS (corr)  \\ \hline
AGN & ELG & 0.108 ( 0.08) & 0.246 (0.21) \\
AGN & Galaxy & 0.005 (0.00) & 0.025 (0.02) \\
AGN & Cluster & 0.077 (0.06) & 0.002 (0.00) \\
AGN & Star & 0.000 (0.00) & 0.000 (0.00) \\
AGN & M Star & 0.004 (0.00) & 0.000 (0.00) \\
ELG & Galaxy & 0.052 (0.03) & 0.115 (0.08) \\
ELG & Cluster & 0.253 (0.17) & 0.079 (0.03) \\
ELG & Star & 0.041 (0.02) & 0.000 (0.00) \\
ELG & M Star & 0.013 (0.01)  & 0.001 (0.00) \\
Galaxy & Cluster & 0.043 (0.01) & 0.267 (0.17) \\
Galaxy & Star & 0.275 (0.13) & 0.268 (0.22) \\
Galaxy & M Star & 0.020 (0.00) & 0.454 (0.22) \\
Cluster & Star & 0.006 (0.01) & 0.000 (0.00) \\
Cluster & M Star & 0.001 (0.00) & 0.004 (0.00) \\
Star & M Star & 0.057 (0.01) & 0.057 (0.01) \\ \hline
\end{tabular}
\end{table}

\section{Model Fitting}\label{sec-fits-model}

\subsection{The fitting technique}

A simple two colour diagram can only provide information in a general sense
about the X-ray emission of the RIXOS sample.  In order to gain a deeper
understanding it is necessary to fit models.  Two main approaches have been
used in obtaining spectral information for similar survey data.  The first is a
simple hardness ratio to determine the power-law slope of low count rate data
and $\chi^2$ fitting for sources with enough counts (e.g. Ciliegi et al. 1996).
However, this approach has the twin disadvantages of not analysing the data in
a uniform way and not taking into account the Poissonian nature of the data for
weak sources.  The other method is to sum up the spectra for sources with
similar $N_H$ and use a standard $\chi^2$ fit to the summed data.  This allows
us to have reasonably high resolution spectra, but has the disadvantage of
losing all information about the individual sources within each $N_H$ band.

\begin{figure}
\psfig{figure=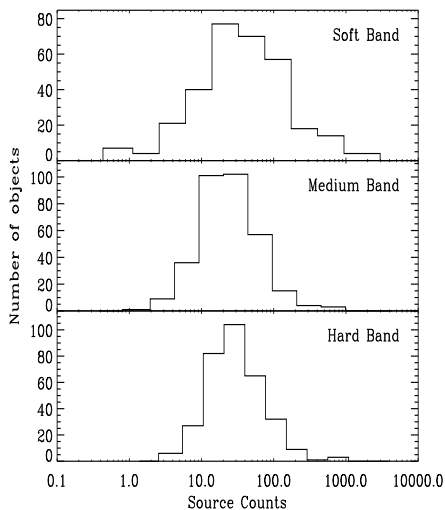,height=8truecm,width=8truecm,angle=0}
\caption{Total source counts in each of the three-colours (soft - 
channels 8-41, medium - channels 52-90, hard - channels 91-201).  Many of the 
sources are close to or at the Poisson limit} 
\label{total_counts}
\end{figure}

We have addressed these problems by fitting two parameter models to our three
colour data for each individual source.  By using three-colours, we can
maximise the signal in each band while retaining one degree of freedom for the
fitting process.  It is also possible to take into account the poissonian
nature of the data directly, by minimising the correct statistic. That there is
a requirement to use such a statistic is clear from Figure~\ref{total_counts}
as many of our sources have $< 15$ counts in one or more of the three
spectral bands used.

A statistic appropriate to the Poisson regime is described by Cash (1979).
This has been successfully applied to the problem, among others, of source
searching in both the WFC all sky survey (Pounds et al. 1993) and the EUVE all
sky survey (Bowyer et al. 1994).  For spectral fitting of low countrate
sources a maximum likelihood method using the Cash statistic is 
appropriate, instead of minimising $\chi^2$ as in the Gaussian regime.

The Cash statistic is derived from the probability of observing $n$ counts
for a given mean $\mu$.  In the Poisson regime this is given by

\begin{equation}
P = \frac{\mu^n e^{-\mu}}{n!}
\end{equation}

therefore for a distribution of counts $n_i$ with predicted means in each
bin $\mu_i$, the total probability is given by

\begin{equation}
P = \prod_{i=1}^{N} \frac{\mu_i^{n_i} e^{-\mu_i}}{n_i!}
\end{equation}

By converting this into a maximum likelihood formulation, we then arrive
at the Cash statistic

\begin{equation}
C = -2 logP = -2 \sum_{i=1}^{N} n_i log(\mu_i) - \mu_i - log(n_i!)
\end{equation}

As $\log(n!)$ is a constant, we can drop it from the calculation since
we are only interested in the minimum of $C$, not its absolute value.  

To fit the data we must arrive at a set of predicted values for $\mu_i$ which
minimise $C$.  To maintain the strict Poissonian nature of the data we
fit the total number of observed counts from the source and background 
within a circle of radius $r_i$
in each band with a mean $\mu_i$ 
i.e. we minimise

\begin{eqnarray}
C' & = & -2 \sum_{i=1}^{N} n_i log(PSF(r_i) \times 
model_i(\alpha_1,\alpha_2,\alpha_3) \nonumber \\ 
   &   & + b_i) - (PSF(r_i) \times model_i(\alpha_1,\alpha_2,\alpha_3) + b_i) 
\label{c:1}
\end{eqnarray}

\noindent where $model_i(\alpha_1,\alpha_2,\alpha_3)$ is the predicted total
counts in band $i$ given some model defined by $\alpha_1,\alpha_2,\alpha_3$.
$PSF(r_i)$ is the fraction of the PSF contained within a radius $r_i$ and $b_i$
is the background contained within radius $r_i$.  For the case of a power-law,
$\alpha_1,\alpha_2$ and $\alpha_3$ would be the normalisation, the power-law
index and the amount of Galactic absorption respectively.  Note that
equation~\ref{c:1} assumes that the background is known to a much higher level
of statistical accuracy than the source counts such that the error on the
background is negligible.  For the RIXOS data, this is the case (see section
3).

Not only does this method deal correctly with the Poissonian nature of the
data, but it also enables us to obtain estimates of the spectrum when we have
upper limits in one or more of the three bands. As the method fits the total
observed counts (source plus background), it automatically takes into account
such upper limits. This is because even in those cases where the predicted
background is larger than the observed number of counts, the predicted number
of source counts [i.e. $model_i(\alpha_1,\alpha_2,\alpha_3)$] will always be
greater than zero.  The case where no source counts are detected is then taken
as a simple statistical fluctuation of the model predicted {\it positive}
source counts.

\subsection{Error Estimation}\label{sec-error}

\begin{figure}
\psfig{figure=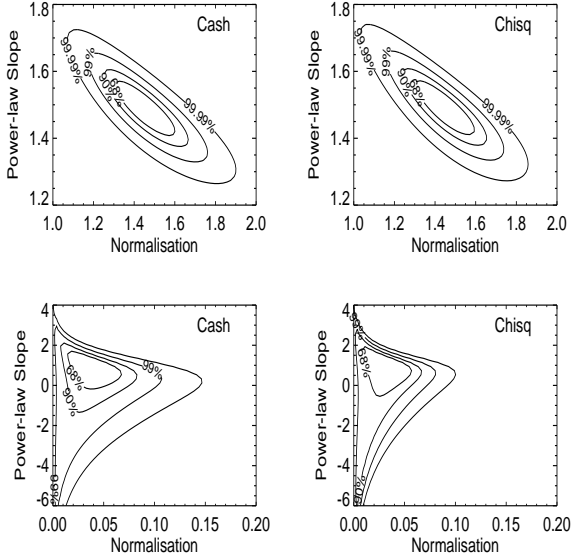,height=8truecm,width=8truecm,angle=0}
\caption{Confidence limits both for a bright source (upper panels) and a faint
source near the Poisson limit (lower panels).  The left hand panels shows the
confidence contours obtained using the Cash statistic, the right had panels
shows the same confidence contours using $\chi^2$.  In the case of the bright
source there is no appreciable difference between the two methods, but for the
faint source the Cash method gives a better constrained slope.} \label{f:1}
\end{figure}

Once we have found a minimum of the Cash statistic, the next step is to
calculate the confidence limits on the fitted parameters.  This can be done in
an identical way to the procedures standardly used in $\chi^2$ fitting, as
the $\Delta C$ statistic is distributed as $\Delta \chi^2$ (Cash 1979).
However, for those sources near the Poisson limit it is difficult to write down
a simple number as the error on a given parameter, because the confidence
contours tend to be asymmetric. Figure~\ref{f:1} show examples of the
confidence contours for both a source near the Poisson limit and a source with
a large number of counts.  In the case of a source near the Poisson limit the
probability contours from a $\Delta \chi^2$ surface and a $C$ surface are
markedly different, with a tighter constraint on the power-law slope from the
$\Delta C$ contours. On the other hand, in the case of a bright source the two
contours are essentially identical.

\begin{figure}
\psfig{figure=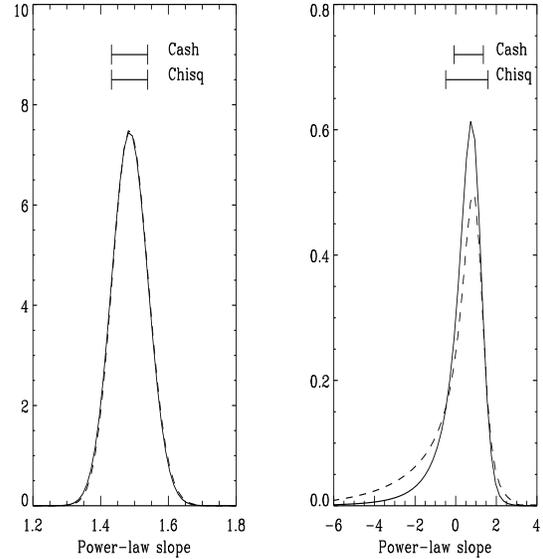,height=8truecm,width=8truecm,angle=0}
\caption{The marginalised $\Delta C$ statistic (which is related to the
probability) as a function of spectral slope for a bright source (left hand
panel) and faint source (right hand panel).  Both the curves for the $\chi^2$
case (dashed) and the Cash statistic case (solid) are shown.  Also shown are
the corresponding marginalised errors for each case.}\label{f:3}
\end{figure}

Owing to the lack of symmetry in the shape of the contours for sources near
the Poisson limit, we have obtained marginalised errors (Loredo 1990 and
references therein). These errors are obtained by integrating the $\Delta C$
values over the unwanted parameters leaving a one dimensional probability for
the parameter of interest.  This then gives us both the most probable value and
the confidence intervals for the parameter of interest in a way that is
statistically independent of any other parameters. The solid lines in
Figure~\ref{f:3} show the probability curves for the power-law slope for both a
weak and a strong source.  In the case of the weak source, the $\chi^2$
probability curve and associated errors are larger than the corresponding Cash
curves. For the bright source they are essentially identical.  This is
precisely the behaviour expected, as the Cash statistic is the same as
$\chi^2$ in the limit of large numbers, and shows the decrease in the size of
the errors bars when the correct statistic is used.

\subsection{Tests of the method \label{cash_test}} 

\begin{figure}
\psfig{figure=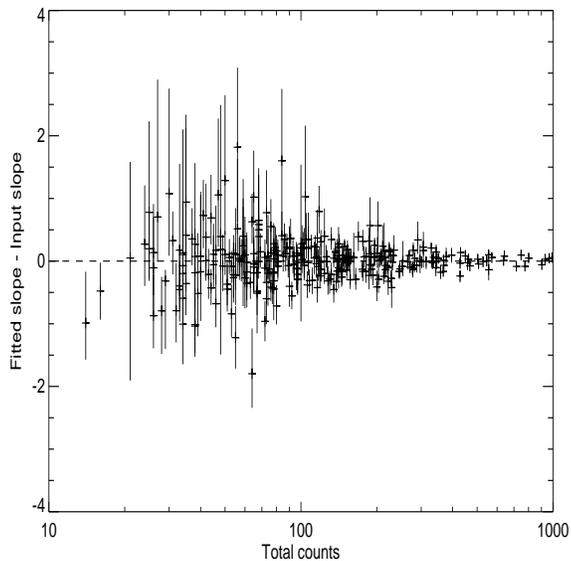,height=8truecm,width=8truecm,angle=0}
\caption{The fitted slope minus input slope for the simulated data.  It is
clear that the Cash method can recover the correct power-law slope over a large
range of source counts.}
\label{diff-slopes}
\end{figure}

\begin{figure}
\psfig{figure=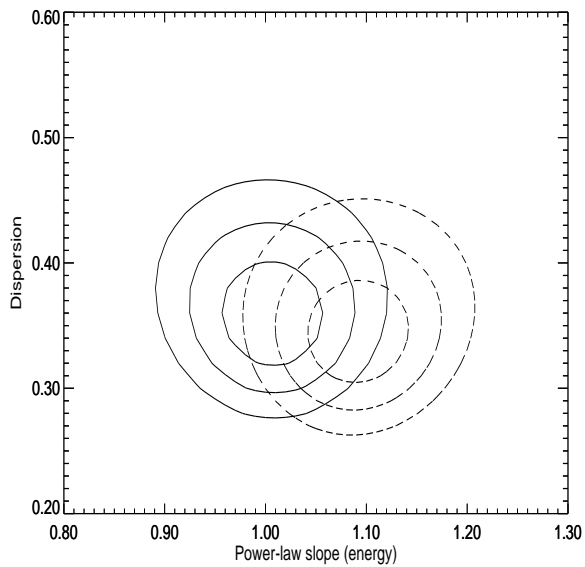,height=8truecm,width=8truecm,angle=0}
\caption{Estimated mean power-law slope and dispersion based on the simulated
data for both the three-colour method (solid line) and the hardness ratio
method (dashed line).  It is clear that the three-colour method gives the same
answer as the input parameters to the simulation, whereas the hardness ratio
method has a bias towards steeper slopes.}
\label{f:4}
\end{figure}

We have adopted a novel approach to the fitting of our data, and to convince
ourselves of their reliability we have run stringent tests.  In particular, we
have investigated the improvement gained by using the correct statistic
relative to using a simple hardness ratio.  In the hardness ratio method the
error is normally derived assuming Gaussian errors of the form $\surd(counts)$,
which in the extreme Poisson limit is no longer strictly true.  In order to
investigate this, we have generated a simulated dataset where each of the
sources has a known input spectrum but the normalisation of the model has been
scaled to give the same number of total counts as each of our real sources.
The individual observed counts in each colour for each source have then been
randomly obtained assuming Poisson statistics. In this way, we have a similar
range of total observed counts and backgrounds to that of our real sample but
with well defined spectral characteristics.  To compare this with the results
for the AGN in the RIXOS sample, the power-law slopes were drawn from a
gaussian distribution of slopes with a mean of $\alpha = 1$ and a dispersion of
$\sigma = 0.4$. The simulated data were then fitted in exactly the same way as
the real data, and the power-law slopes and errors were determined from the
marginalised errors.  Figure~\ref{diff-slopes} shows the fitted slope minus the
input slope for each source as a function of the source counts, and shows that
the Cash method can recover the correct slope over a large flux range.
Further, from our fitted slopes and errors we have estimated the average
power-law slope and dispersion using the method outlined in Nandra and Pounds
(1994) and Maccacaro et al. (1988).  However, instead of assuming gaussian
statistics when dealing with the errors, we use the probability curves derived
from the $\Delta C$ surfaces. The confidence limits of the mean power-law slope
and intrinsic dispersion from the simulated data are shown in Figure~\ref{f:4}.
The results are in excellent agreement with the input values giving $\alpha =
1.02 \pm 0.05$ and $\sigma = 0.36 \pm 0.05$.  We have also analysed the same
dataset using a hardness ratio method, where we have used the ratio $S/(H1+H2)$
to estimate the spectral slope together with an error based on gaussian
statistics.  We have determined the average power-law slope and intrinsic
dispersion of the sources from the hardness ratios and the result is shown by
the dashed contours. A comparison between the result obtain by using the Cash
statistic and the hardness ratio method shows that the hardness ratio result is
marginaly biased towards steeper (softer) slopes.  It is likely that some bias
may be caused by the failure of the hardness ratio methods to take into account
the Poissonian nature of the data, so this effect will depend on how many faint
sources (i.e. with few counts) are contained within any given sample.
Therefore, it is possible that the use of a hardness ratio may have caused some
bias in the results of previous surveys.

\section{Model fits to the RIXOS data} \label{sec-fits}

As we have limited resolution with three-colours, our initial model is a
simple power-law.  For each extra-galactic source we have fixed the value of
the Galactic $N_H$ at the Stark et~al. (1992) HI value and left the
normalisation and slope as free parameters.  For the stars we have simply set
the $N_H$ at zero.  Each source is fitted using the relevant response matrix
for the source position and date of observation.  The assumption of a power-law
fit to all classes of objects is clearly incorrect for many of the sources, for
example the stars and clusters, so these fits are only indicative of the
overall slope of the X-ray spectrum.  However, for the AGN it is likely to be
fairly representative of the true flux distribution from many of the sources.

\begin{figure}
\psfig{figure=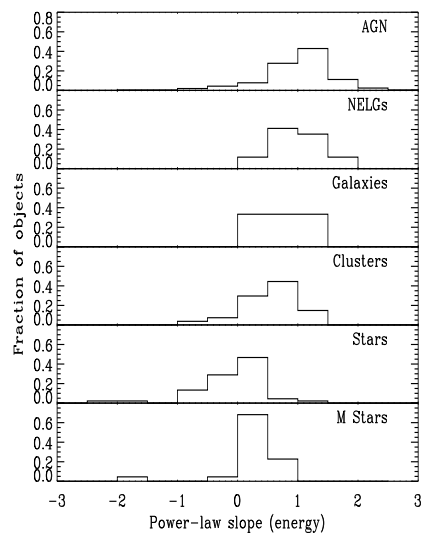,height=8truecm,width=8truecm,angle=0}
\caption{Histograms showing the distribution 
of the fitted power-laws divided into different object
classes.} \label{fig-hist}
\end{figure}

All of the fitted slopes are listed in Table 2. They have been determined in
two different ways.  Column 14 quotes the marginalised slope and error derived
in a way which is independent of the value of the normalisation (see
Section~\ref{sec-error} for details).  Columns 15 and 16 of Table 2 list the
normalisation and slope derived from the minimum on the Cash surface for each
source.  As the marginalised slopes are independent of the other parameters, we
have used this value in all subsequent plots rather than the value derived from
the minimum of the Cash surface. The flux derived from the fits is given in
column 17 where the error on the flux has been obtained by folding the flux
calculation through the Cash contour.  This flux may differ from the flux used
to establish the sample since the fitted slope may be significantly different
from a slope of 1.  Further, for uniformity all sources have been treated as
point like in the present analysis and no attempt has been made to correct for
extended sources, which was done in deriving the original fluxes.
Figure~\ref{fig-hist} shows the distribution of slopes for each class of
objects.  Essentially, the distribution of slopes again indicates differences
between the AGN/NELGs and the clusters and stars. However, with the fitted
data, we can investigate the intrinsic spectrum of the individual classes of
sources in more detail.

\subsection{The Stars}

\begin{figure}
\psfig{figure=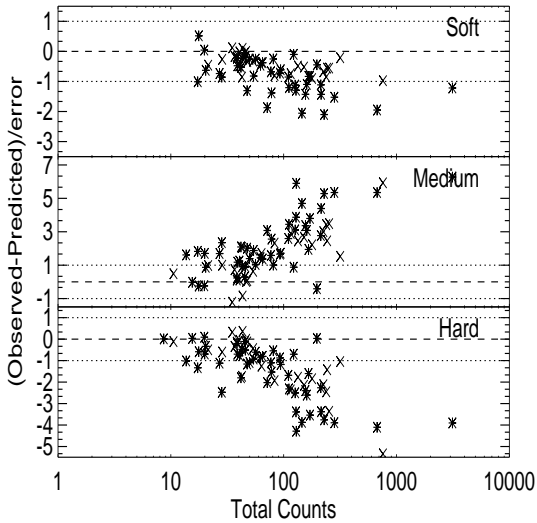,height=8truecm,width=8truecm,angle=0}
\caption{The predicted minus observed counts relative to a power-law model 
for the stars. The dotted lines denote 1 sigma around zero.
The different symbols denote the two different types of stars
in the RIXOS sample ($\star$ stars and $\times$ dMe stars)} \label{fig-stars}
\end{figure}

The average power-law slope for the stars is $\alpha = 0.40$, implying that as
a class they are hard sources. However, a power-law fit to the data is unlikely
to be a good representation of the stellar X-ray emission.  Though no simple
way exists to determine the goodness of fit directly from the Cash statistic,
it is possible to distinguish between good and bad fits to the data.  To do
this, we have calculated the expected number of counts based on the fitted
model, subtracted the actual observed counts, and divided by  the square root
of the observed counts (an estimate of the error on the source counts).
Figure~\ref{fig-stars} shows this quantity  for the best fit power-law model
for each of the three-colours for each star. It is clear from this plot that a
power-law fit is not a good model of the stellar X-ray emission.  It 
consistently over estimates the S1 and H2 bands, while consistently
under estimating the H1 band.  This distribution of the data relative to a
power-law fit is, however, entirely consistent with the emission arising from
warm ($< 3$ keV) gas.  Such temperatures give rise to a large amount of line
emission, particularly around the iron complex at 1 keV, and this line emission
is the most likely explanation for the deviations seen in the three-colour
data, particularly in the medium band.

\begin{figure}
\psfig{figure=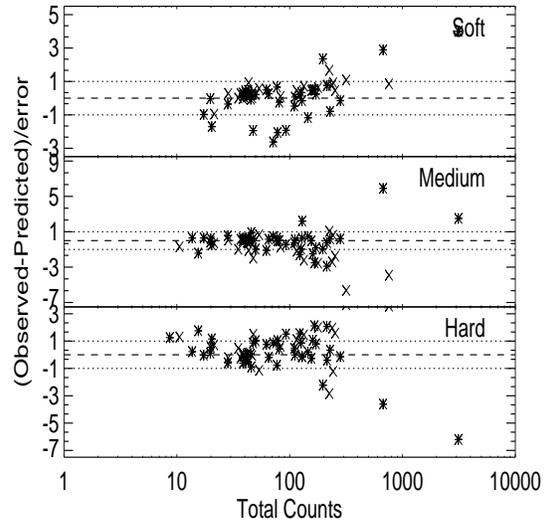,height=8truecm,width=8truecm,angle=0}
\caption{The Predicted minus observed counts relative to a Raymond and Smith
model for the stars.  The dotted lines denote $1\sigma$ around zero.
The use of a Raymond and Smith model has improved the 
goodness of fit for many of the stars.} \label{fig-stars2}
\end{figure}

With only three-colour data, it is not possible to fit the multi-temperature
models known to be required for X-ray spectra of stars (e.g. Schmitt et
al. 1990).  We have, however, fitted a single temperature Raymond and Smith
model to our data. Figure~\ref{fig-stars2} shows the predicted minus observed
counts with respect to the Raymond and Smith fits for the stars. It is clear
that there is a marked improvement over the power-law fits, demonstrating that
three-colour data are capable of distinguishing between thermal and non-thermal
models.  

\subsection{The Clusters}

\begin{figure}
\psfig{figure=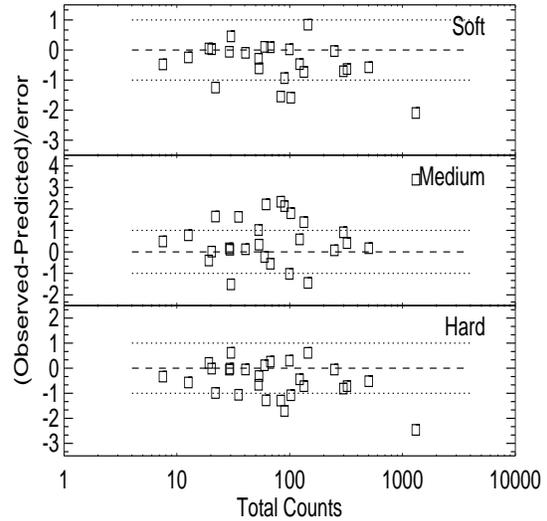,height=8truecm,width=8truecm,angle=0}
\caption{The predicted minus observed counts for the clusters.
The dotted lines denote $1\sigma$ around zero.} 
\label{fig-clusters}
\end{figure}

Figure~\ref{fig-clusters} shows the predicted minus observed distribution for
each of the three bands for power-law fits to the cluster data.  From this, it
is apparent that a power-law fit is a reasonable representation of the data for
most, though not all, clusters.  At face value, this may seem surprising, since
it is known that cluster emission arises from hot ($2 - 10$ keV) intergalactic
gas with some clusters showing evidence for a cooling flow (e.g. Sarazin 1986).
However, because of the low energy of the ROSAT passband ($0.1 - 2$ keV), a hot
plasma spectrum ($> 3$ keV) is fairly well modelled by a power-law with a slope
$\alpha \sim 0.5$.  This power-law slope is relatively insensitive to
temperature and $N_H$.  If we determine the average slope for the clusters, we
find a mean of $\alpha = 0.5 \pm 0.05$, which is in agreement with that
expected for a bremsstrahlung model with a temperature $kT > 3$ keV.  There
are, however, a number of clusters which, like the stars, show deviations from
a simple power-law, and it is likely that these objects have temperatures lower
than $3$ keV.

\begin{figure}
\psfig{figure=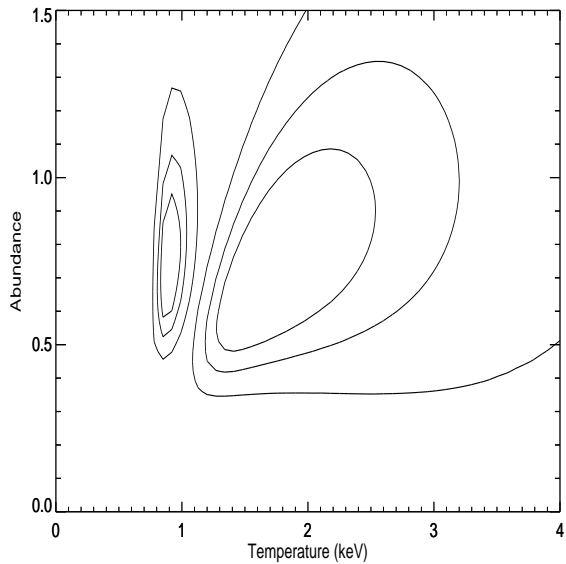,height=8truecm,width=8truecm,angle=0}
\caption{Colour plot of the Raymond and Smith fit to 240-564 in two regions,
one in the inner most arc-minute, the other from an annular region
surrounding the central core.  There is a clear cooling of the temperature
between the inner regions and the outer regions.}
\label{fit-cluster1}
\end{figure}

As the use of a single temperature Raymond and Smith model reduced the
residuals to the fit for the stars, we have attempted to fit a similar model to
the three-colour data for the clusters.  However, unlike the stars we find that
in some cases a single temperature model does not reduce the residuals.  This
was particularly true of 240-564, the brightest source in our sample, where a
single temperature fit to the three-colour data gave a high ($> 10$ keV)
temperature.  Based on the residuals to a power-law fit, 240-564 would be
expected to have a relatively low temperature.  For this object we have been
able to extract a high resolution spectrum which we have fitted using XSPEC.
Figure~\ref{fit-cluster1} shows the fits both to the central region of the
cluster and to an annular region surrounding the centre.  As expected from the
residuals, the temperature is low.  There is also an apparent reduction in
temperature towards the centre, implying that this cluster at least has a
cooling flow.  It is therefore clear that since three-colour data gave a {\sl
high} temperature we cannot use the temperatures derived for the clusters with
any degree of certainty. We have extended the method to four or five colours on
a number of the clusters within the RIXOS sample and this work shows that, with
the extra channels, the method gives results that are consistent with higher
resolution data. Such an extension to more colours is beyond the scope of this
paper, however.

\subsection{The Narrow Emission Line Objects\label{nelg_sec}}

A subject of great interest is the X-ray emission from narrow emission line
galaxies.  Studies of the $\log N - \log S$ in fainter surveys such as the UK
Deep survey show that the fraction of quasars at very faint flux levels
declines, but that of NELGs rapidly increases (McHardy et
al. 1998). Extrapolations to zero flux indicate that up to 50\% of the X-ray
background may be due to NELGs.  The spectral shape of NELGs is therefore of
crucial importance if we are to understand the nature of the soft X-ray
background.  However, a fundamental problem with this class of object is that
the term NELG is a nebulous categorisation.  They include hidden AGN, such as
Seyfert 2's where the emission is likely to be non-thermal and absorbed, to
starburst galaxies and HII region galaxies, where the emission is thought to be
thermal in nature and arising from shocked gas with a typical temperature of
$0.5 - 1$ keV.  From our work on stars, we know that we can distinguish between
thermal and non-thermal sources on the basis of the fits to the three-colour
data, so we should be able to estimate the fraction of thermal to non-thermal
sources in the RIXOS sample.

\begin{figure}
\psfig{figure=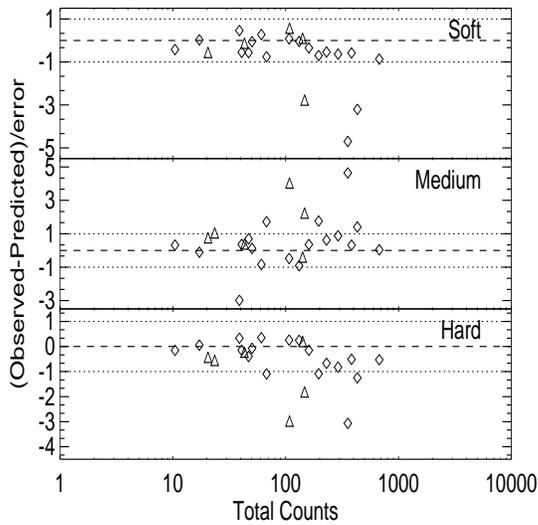,height=8truecm,width=8truecm,angle=0}
\caption{The predicted minus observed counts for the NELGs ($\diamond$)
and isolated galaxies ($\bigtriangleup$). The dotted lines denote $1\sigma$ around zero.} 
\label{fig-nelg}
\end{figure}

\begin{figure}
\psfig{figure=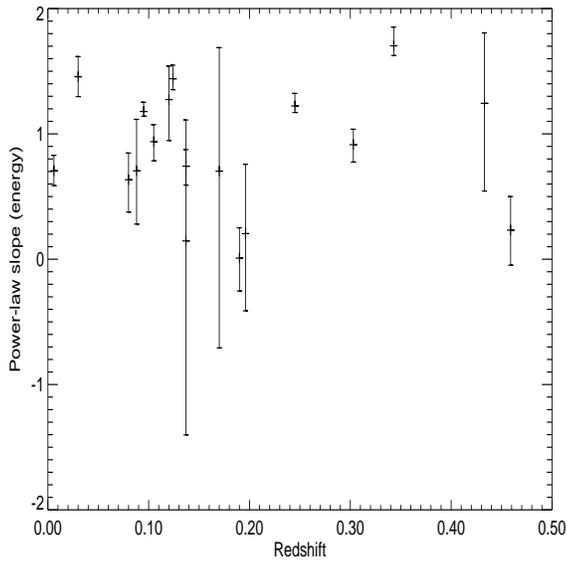,height=8truecm,width=8truecm,angle=0}
\caption{The fitted power-law slope versus the redshift for the NELGs in the
RIXOS sample.  No evolution of the power-law slope can be seen.} 
\label{fit-nelg_z}
\end{figure}

There are 18 NELGs in RIXOS identified on the basis of their optical spectra.
Figure~\ref{fig-nelg} shows the distribution of observed minus predicted counts
for the NELGs.  In general, the NELGs seem consistent with a power-law, with
only one source (122-16) showing a significant deviation in all three bands.
This may imply thermal emission from this object, though preliminary studies of
higher resolution data from 122-16 indicates that the X-ray spectrum is
complex.  Figure~\ref{rixos_slopes} shows the estimated mean slope and
dispersion of the NELGs in comparison with the AGN and figure~\ref{fit-nelg_z}
shows the power-law slope as a function of redshift. This demonstrates no clear
evidence for spectral evolution with redshift.  Figure~\ref{fit-nelg_z} does
show that there is a large range of potential slopes, with some of the sources
being very hard.  At least one of these hard sources has been identified with a 
Seyfert 2 galaxy, which is entirely consistent with the flat spectral slope.

Thus the X-ray spectra of NELGs in the RIXOS sample are indistinguishable 
from those of the AGN. This is consistent with the fact that high resolution
optical data on X-ray selected NELGs has shown that many objects classified as
NELGs on the basis of low signal-to-noise data have broad components to the
permitted lines (e.g. Boyle et al. 1995), and at least two H II region like
galaxies have been observed to show strong X-ray variability more consistent
with that seen in AGN (Boller et al. 1994, Bade et al. 1996).  Further, HRI
images of low-luminosity AGN show that the X-ray emission is mostly nuclear,
again supporting the idea that the origin of the X-ray emission is nuclear in
nature (Koratkar et al. 1995).  Thus  many objects classified as NELGs above
the RIXOS flux limit may contain active nuclei.

However, it is important to note that the average slope that we find for  the
RIXOS NELGs is inconsistent with the average slope of a sample of much fainter
NELGs from the UK Deep Field (Romero-Colmenero et al. 1996).  In the latter
study the average slope was $\alpha = 0.45 \pm 0.09$, a value that is more
consistent with the average slope of the RIXOS clusters.  There is clearly a
discrepancy between the average properties of NELGs in the RIXOS sample and
those found at much fainter fluxes which may imply some difference in the type
of objects seen at the faintest fluxes.  Another possibility is that there are
more absorbed sources in the fainter samples which would pull down the average
slope. Without higher resolution data and good signal to noise it is impossible
to distinguish between these two possibilities.

\subsection{The AGN}

\begin{figure}
\psfig{figure=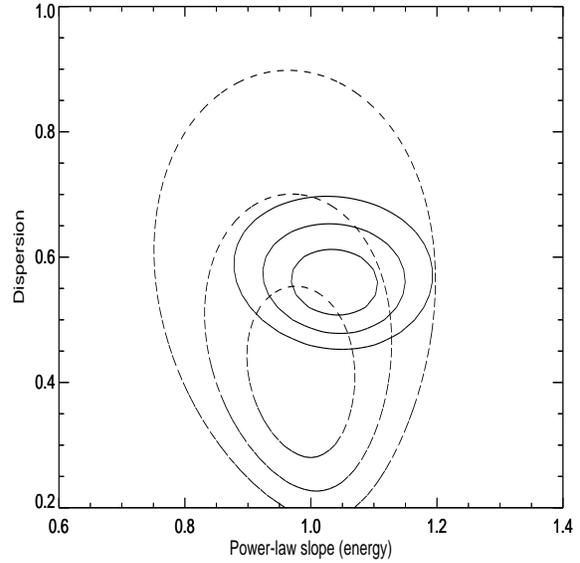,height=8truecm,width=8truecm,angle=0}
\caption{The values of the slope and dispersion for the RIXOS AGN (solid
line) and the NELGs (dashed line).  It is clear that the two samples are
indistinguishable.}\label{rixos_slopes}
\end{figure}

By far the largest fraction of objects in the RIXOS sample have been classified
as AGN.  Unlike the stars and clusters of galaxies, a non-thermal model such as
a power-law is likely to be an acceptable fit to the data, though in detail
more complex models may be appropriate (e.g. Nandra and Pounds 1994). 
Figure~\ref{rixos_slopes} shows the mean slope and dispersion for the RIXOS
AGN; we find $\alpha = 1.05 \pm 0.05$ and an intrinsic scatter of $0.55 \pm
0.05$.  These numbers are slightly steeper than those found for the EMSS AGN
($0.9 \pm 0.05$ with a dispersion of $0.35 \pm 0.04$), which sampled a harder
energy range ($0.3 - 4$ keV) than the PSPC.  However, this is much flatter than
the slopes found for bright, nearby Seyferts observed with ROSAT (e.g. Walter
and Fink 1993, Laor et al. 1994, Fiore et al. 1994) and is also flatter than
the average slope found for the fainter AGN contained in the CCRS ($\alpha =
1.3 \pm 0.1$ Ciliegi et al.  1996).  A slope of $\alpha = 1$ is, however,
consistent with what is believed to be the underlying power-law slope in nearby
Seyferts (e.g. Nandra and Pounds 1994).

\subsubsection{Goodness of fit for a power-law}

\begin{figure}
\psfig{figure=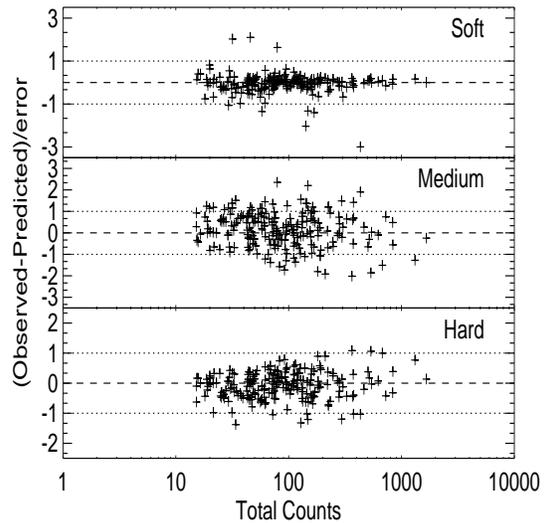,height=8truecm,width=8truecm,angle=0}
\caption{The Predicted minus observed counts for the AGN. 
The dotted lines denote 1 sigma around zero.} \label{fig-AGN}
\end{figure}

Figure~\ref{fig-AGN} shows the predicted minus observed total counts expressed
in terms of the standard deviation for each spectral band for all the AGN.  It
is clear that for the majority of AGN a power-law fit is a reasonable
representation of the spectral shape, as most of the data points lie within
one sigma of the model. However, there do seem to be a number of AGN where the
observed counts are significantly under estimated in the medium band.  It is
not clear what causes this deviation, as these sources seem to contain a
mixture of slopes ranging from soft to hard.  One possibility is that some
contain a significant O~VII edge, implying the presence of a warm absorber.
However, three-colour data are not sufficient to determine the origin of this
deviation and higher resolution data are required.  We have extracted high
resolution data for the source with the largest discrepancy in the soft band,
source 258-001, which is sufficiently bright to warrant this.  Analysis of
these data shows evidence for an edge at 1.1 keV which has been tentatively
identified with silicon (Mittaz et al. 1998).  The ability to detect such a
source shows the power of fitting three-colours to reveal peculiar
features. From Figure~\ref{fig-AGN} approximately 20\% of sources appear to be
deviant from a power-law model.

\begin{figure}
\psfig{figure=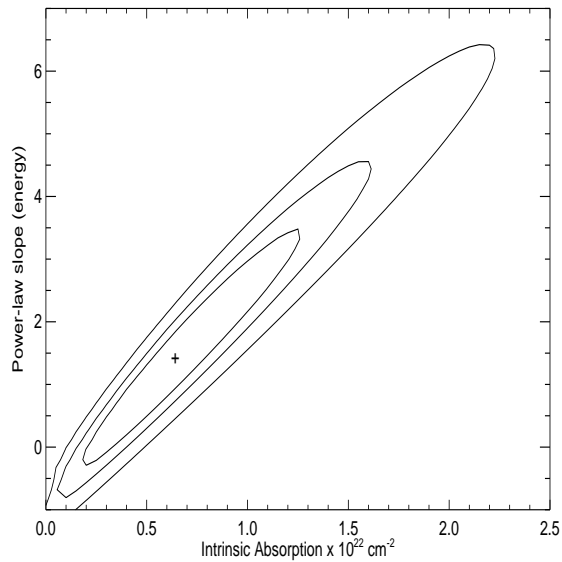,height=8truecm,width=8truecm,angle=0}
\caption{The contour plot of power-law slope against $N_H$ for 278-010.
Intrinsic  absorption can be clearly seen} \label{fig3}
\end{figure}

Without analysing all of the data at higher resolution, it is difficult to make
strong claims about objects where a simple power-law does not appear to be an
adequate description of the data.  As noted in section~\ref{colour-colour-sec},
within the sample there are a number of sources which significantly deviate
from the average spectrum.  For example, some AGN have positive slopes, which
correspond to those in the $C1 < 0$ region of the colour colour diagram. Such
objects may be intrinsically absorbed and one (278-010) has sufficient counts
to allow us to extract a higher resolution spectrum.  On the assumption that
the absorbing column is at the Stark et~al. (1992) value of $1.94 \times
10^{20}$ cm$^{2}$ a fit to these higher resolution data gives a slope of
$\alpha = -1.17 \pm 0.2$, consistent with the value fitted to the three-colour
data.  However, if we fit an intrinsic column in addition to the galactic
$N_H$, we detect an instrinsic column at $>90$\% confidence with a fitted
power-law slope of $\alpha = (1.4^{+2.3}_{-1.7})$ and a best fit intrinsic
$N_H$ of $(6^{+6}_{-4}) \times 10^{21}$ cm$^{-2}$ (68\% confidence limits).
Figure~\ref{fig3} shows the 68\%, 90\% and 99\% contours of the intrinsic
absorption plotted against power-law slope.  On the assumption that this hold
true for the other AGN with $C1 < 0$ we can conclude that $\sim$5\% of the
RIXOS AGN sample show detectable amounts of intrinsic absorption.  We note that
trends between the fitted X-ray and optical spectral slopes, and between the
X-ray spectral slope and the ratio of X-ray to optical flux, of RIXOS AGN have
also be interpreted as being due to the effects of absorption (Puchnarewicz
et~al.  1996).

\begin{figure}
\psfig{figure=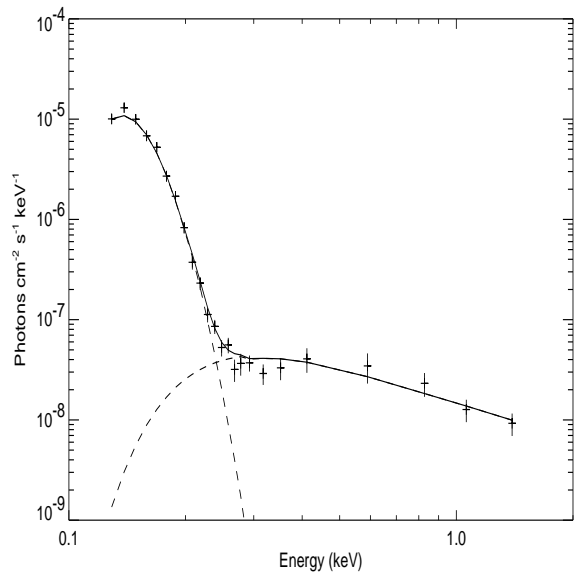,height=8truecm,width=8truecm,angle=0}
\caption{A multicomponent fit to the X-ray spectrum of 227-301.
The best fit model (black body and power-law) is shown} \label{fig-scont}
\end{figure}

From the fitted power-law slopes it is moreover clear that there are not only
hard sources, but also those which have slopes significantly steeper than
$\alpha = 1$.  We have taken the object with the steepest slope which has a
sufficiently large number of counts (227-301) and have extracted higher
resolution X-ray data for it. A single power-law gives a very bad fit, with a
$\chi^2_\nu$ of 6.8.  We therefore fitted the data with a power-law and
black-body model (to represent any soft excess), and the fit improved
dramatically, with a $\chi^2_\nu$ of 0.56.  The best fit parameters give a
power-law slope of $\alpha = 0.7 \pm 0.7$ (68\%) and a black-body temperature
of $0.0085 \pm 0.001$ keV (68\%) (Figure~\ref{fig-scont}).  The value of the
power-law slope is now consistent with the average for the RIXOS AGN, and the
black-body component has a similar temperature to that seen the USS sample
(Thompson and Cordova 1994).  It is therefore clear that RIXOS contains a range
of objects, from intrinsically absorbed AGN and those with strong soft excesses
to objects with absorption edges.  The high resolution data show that fits to
the three-colour data can give sufficient information to separate out those
objects which have non-standard X-ray spectra.


\subsubsection{Spectral Evolution}

\begin{figure*}
\psfig{figure=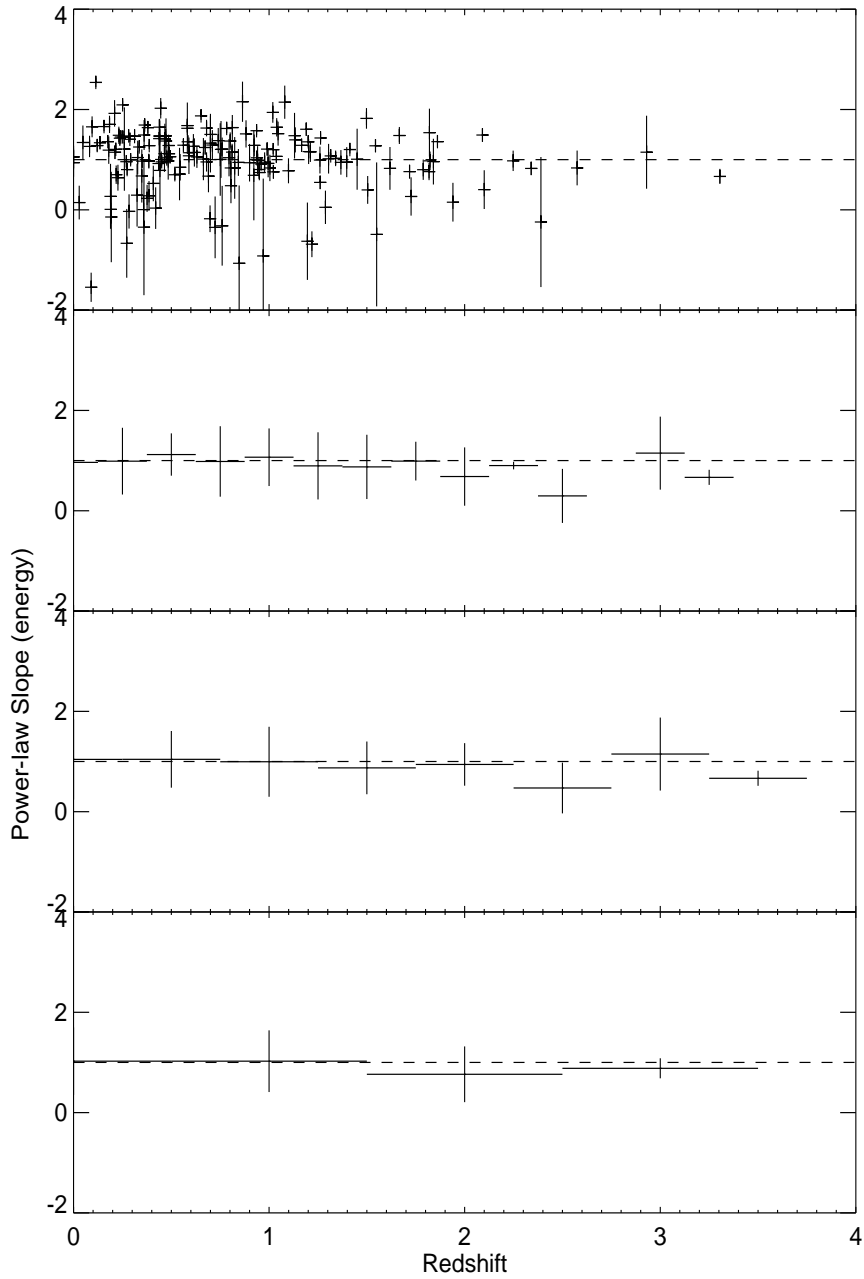,height=20truecm,width=13truecm,angle=0}
\caption{Fitted slopes for the RIXOS AGN as a function of redshift.  Each panel
shows the data binned into successively bigger redshift bins with the error bar
representing the RMS scatter about the mean, and shows that there is no
evolution in the spectral slope.  In the case of the middle two panels, the
last two data points contain a single object and the plotted error is simply
the error on the fit to those objects.
\label{evolAGN}}
\end{figure*}

\begin{figure}
\psfig{figure=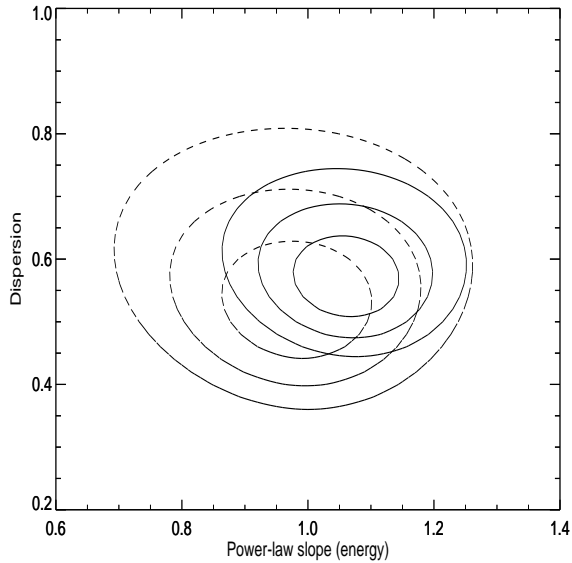,height=8truecm,width=8truecm,angle=0}
\caption{Intrinsic slope and dispersion for the RIXOS AGN separated into
objects above and below a redshift of 1.  It is clear that the two samples are
identical implying no strong evolution in the spectral parameters of the AGN.
\label{slopes_evol}}
\end{figure}

A further question of interest is whether there is any evolution of the  X-ray
spectral slope of AGN with redshift.   The nature of any such spectral
evolution has consequences for both our understanding of the X-ray emitting
process and the nature of the X-ray background.  Figure~\ref{evolAGN} shows the
distribution of slopes as a function of redshift, where we have used variously
sized redshift bins.  It is apparent from Figure~\ref{evolAGN} that there is no
evidence for any spectral evolution at all.  This lack of evolution in the
spectral slopes is consistent with the results of other similar samples (e.g.
Ciliegi et al. 1996).  To investigate this further, we have recalculated the
intrinsic slope and dispersion for AGN below and above a redshift of 1, where
we have approximately equal numbers of AGN in each of the two bins.  From
figure~\ref{slopes_evol} it is clear that the slopes and dispersions are
effectively identical for objects above and below the redshift divide. As noted
by Ciliegi et al. (1996), the fact that there is no apparent spectral evolution
implies that the power-law spectrum in the AGN rest frame extends from soft
X-rays out to at least 8 keV with the same slope. This excludes models
with strong or hot soft excesses as being typical of AGN in the RIXOS sample.

\subsubsection{Interpretation and comparison with other surveys}

The standard model for the X-ray emission from AGN derived from missions
previous to GINGA and ROSAT was one of a medium energy power-law with a slope
of $\alpha = 0.7$ with many objects showing evidence for a soft X-ray excess.
Such an excess is normally assumed to be due to the high energy tail of an
accretion disk spectrum (e.g. Turner and Pounds 1988).  However, with the
advent of ROSAT and more recently ASCA, the situation has been found to be more
complex.  In detail it is often necessary to use models including reflection
and warm absorbers as well as simple power-laws (e.g. Nandra and Pounds 1994).
Approximately 50\% of nearby Seyferts studied by Ginga have shown some evidence
for warm absorbers, and evidence for an absorption line at 0.7 keV identified
as O VII has even been found in PSPC data alone (e.g. Nandra and Pounds 1992).


Over the past decade there has been a lot of work on soft X-ray surveys of AGN
and a number of samples have been compiled.  The largest of these is the EMSS
(Gioia et al. 1991) which consists of 421 AGN detected in the $0.5 - 4.5$ keV
band.  Maccacaro et al. (1988) found a mean spectral index for the AGN of
$\alpha = 1.03 \pm 0.05$ with a dispersion of $\sigma = 0.36$. Later surveys
indicated that there may be an average steepening of the spectrum towards
softer energies.  The Ultra-Soft Survey from Einstein showed that AGN selected
below 0.5 keV have an average slope of $\alpha = 1.45$ (Thompson and Cordova
1994).  The EXOSAT High Galactic Survey (HGLS), which covered the energy range
0.2 - 2 keV, was consistent with a mean spectral slope of $\alpha = 1.3$
(Giommi et al.  1988) and work on ROSAT PSPC observations have indicated that
the average spectrum of nearby bright AGN is about $\alpha = 1.5$ (e.g.  Walter
and Fink 1993).  Recently Schartel et al. (1996) used the ROSAT all-sky survey
data for the QSOs in the Large Bright QSO survey and found a mean energy index
of $\alpha = 1.70 \pm 0.2$ for the radio-quiet QSOs.  Moving to higher redshift
samples, Bechtold et al. (1994) found $\alpha = 1.15 \pm 0.14$ for a sample of
high redshift, radio-quiet quasars and Reimers et al. (1995) found $\alpha =
1.25 \pm 0.2$ for another high redshift sample.  It is difficult to make
general statements on the basis of these different samples, however, since each
has its own selection criteria and may therefore sample different populations
of sources.

None of the above samples is directly comparable to RIXOS, either because of
they are optically selected, or select bright X-ray sources, both of which can
favour AGN with soft X-ray spectra (Puchnarewicz et~al. 1996). There are,
however, ROSAT serendipitous surveys with which a direct comparison should be
more meaningful, although none are as large and/or complete as RIXOS.  The
CCRS, which has a similar flux limit to RIXOS, has a reported average slope
which lies between the EMSS average and the average for the brighter samples,
with $\alpha = 1.3 \pm 0.1$ (Ciliegi et al.  1996).  At the very faintest
fluxes, the average spectral slopes for QSOs in the UK Deep survey is $\alpha =
0.96 \pm 0.03$ (Romero-Colmenero et al. 1996) which is consistent with the
EMSS.  The spectrum of the QSOs in another deep survey has an average of
$\alpha = 1.23 \pm 0.04$ (Almaini et al., 1996).  Both the Ciliegi and the
Almaini samples are therefore softer than the RIXOS average of $\alpha = 1.05
\pm 0.05$.

\begin{figure}
\psfig{figure=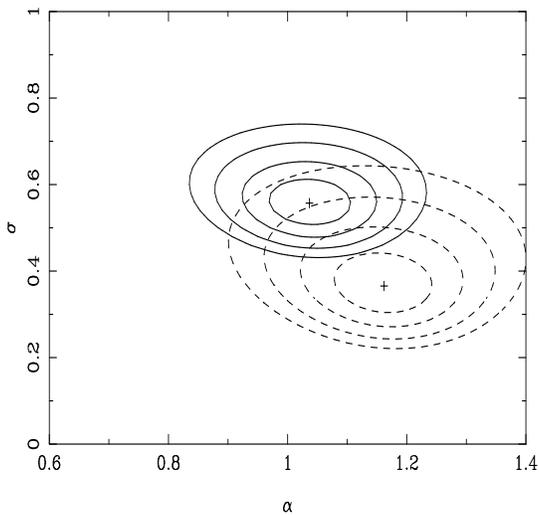,height=8truecm,width=8truecm,angle=-90}
\caption{A comparison of the RIXOS average slope and dispersion for the AGN
(solid contour) and the estimated spectral slope and dispersion for the CCRS
(dashed contour).  Both contours have been obtained in precisely the same way,
using the Cash statistic}
\label{ccrs_rixos}
\end{figure}

The largest difference is between the CCRS sample and RIXOS.  Although the
discrepancy is only at the two sigma level, it would be expected that these two
samples would give essentially identical results, as the flux limits for the
CCRS are only slightly lower than for RIXOS.  However, the analysis techniques
are different. From the simulations described in Section~\ref{cash_test} it
would be expected that the hardness ratio method used in the CCRS would give an
average slope that was slightly softer than the `true' value. We have attempted
to re-analyse the CCRS data using the method used for the RIXOS sample.
However, because accurate positions are not quoted for sources in the CCRS, it
has not been possible to analyse the CCRS in exactly the same way as for RIXOS
since we cannot unambiguously identify all the X-ray sources. Nevertheless
Figure~\ref{ccrs_rixos} shows the comparison between the RIXOS sample and our
best estimate for the CCRS sample analysed using the Cash method, and the
discrepancy between the two is reduced.  The revised average slope of the CCRS,
$1.16 \pm 0.1$, is now consistent with the RIXOS sample at $1\sigma$ implying
that the apparent difference between the two samples was caused at least in
part by the bias introduced by using hardness ratios.  The difference between
the two dispersion estimates is likely to be caused by the assumption of a
gaussian distribution of slopes rather than necessarily representing a real
difference between the two samples.


From the range of different slopes obtained from different samples it is clear
that the spectral distribution of AGN is quite complex and that source
selection effects can play a dominant role in determining the average slope
within a given sample.  Nevertheless, the evidence increasingly suggests that
faint X-ray selected  AGN such as those found in ROSAT serendipitous surveys
have a mean slope close to $\alpha = 1$.  Such an index is close to the value
estimated to be the underlying intrinsic X-ray spectrum in Seyferts when
effects such as reflection and a warm absorber are taken into account (Nandra
and Pounds 1994). It is also close to the value of the inferred spectral slope
seen in the IR, giving rise to claims that there is a power-law of energy index
1 underlying the observed spectrum from the IR to the X-ray range (e.g.~Elvis
et al. 1986). Deviations from this average slope can then be caused by
additional processes such as soft X-ray excesses, warm absorbers and
reflection, the effects of which are likely to be a function of redshift and/or
luminosity. Some of these additional effects have already been observed in some
high redshift objects. For example warm absorbers have been detected so far in
two quasars, 3C351 (Fiore et al. 1993) and MR2251-178 (Pan, Stewart and Pounds
1990) though it is unclear how prevalent they are.  However, without a detailed
study of objects contained within RIXOS and other similar samples it is not
possible to determine the proportion of objects in which these extra effects
are important.

\begin{figure}
\psfig{figure=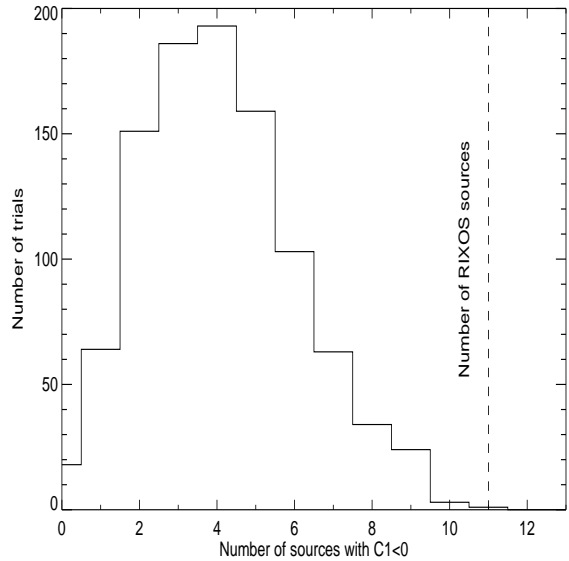,height=8truecm,width=8truecm,angle=0}
\caption{A histogram showing the number of times a given number of hard ($C1 <
0$) sources were found in simulated samples of RIXOS AGN.  The simulated
samples have the same flux and spectral distribution as the RIXOS AGN with a
mean $\alpha=1$ and dispersion $\sigma = 0.55$.  Only one of the simulated 
datasets have as many hard sources as were seen the the RIXOS sample of AGN.}
\label{hard_sim}
\end{figure}

One final question needs to be addressed, and that is the effect of intrinsic
absorption.  Observations of selected high redshift quasars have indicated that
absorption may be important in some objects (Elvis et al. 1994).  High
resolution spectra of CCRS AGN with sufficient counts to determine $N_H$ show
only one object out of 36 AGN with evidence for significant absorption (Ciliegi
et al. 1996), while in the data of Almaini et al. (1996) two out of nine
objects require extra absorption.  Selection effects may be at work here, since
a source that is absorbed will appear fainter than the same source that is not,
and the constraints on spectral fits are obviously better for brighter sources.
However, from the X-ray colour-colour data there are indications that at least
$\sim$ 5\% of the RIXOS AGN have detectable absorption based on the X-ray data
alone. This fraction is fairly secure because even at the flux cut-off of $3
\times 10^{-14}$ ergs cm$^{-2}$ s$^{-1}$ RIXOS sources contain significant
numbers counts. This is confirmed by simulations. We have constructed 1000
datasets with the same flux distribution and assumed slope distribution
($\alpha = 1$ and $\sigma = 0.55$) as the RIXOS sample and have analysed these
datasets in exactly the same way as discussed in
section~\ref{colour-colour-sec}.  Out of the 1000 simulated datasets, only one
of the simulated datasets have as many AGN in the $C1 < 0$ region of the
colour-colour plot as are actually seen, as illustrated in
Figure~\ref{hard_sim}.

Even in those
samples with much lower flux limits than exist in the RIXOS samples there is
no evidence for extra absorption being required for the majority of the QSO
part of the sample.  Therefore, based on the X-ray data alone, there is no
strong evidence for absorption playing a major role in the X-ray spectra of
faint AGN and such effects only exist at the $\sim 10$\% level.



\section{Analysis of the whole RIXOS sample \label{sec-flux_slopes}}

\begin{figure*}
\psfig{figure=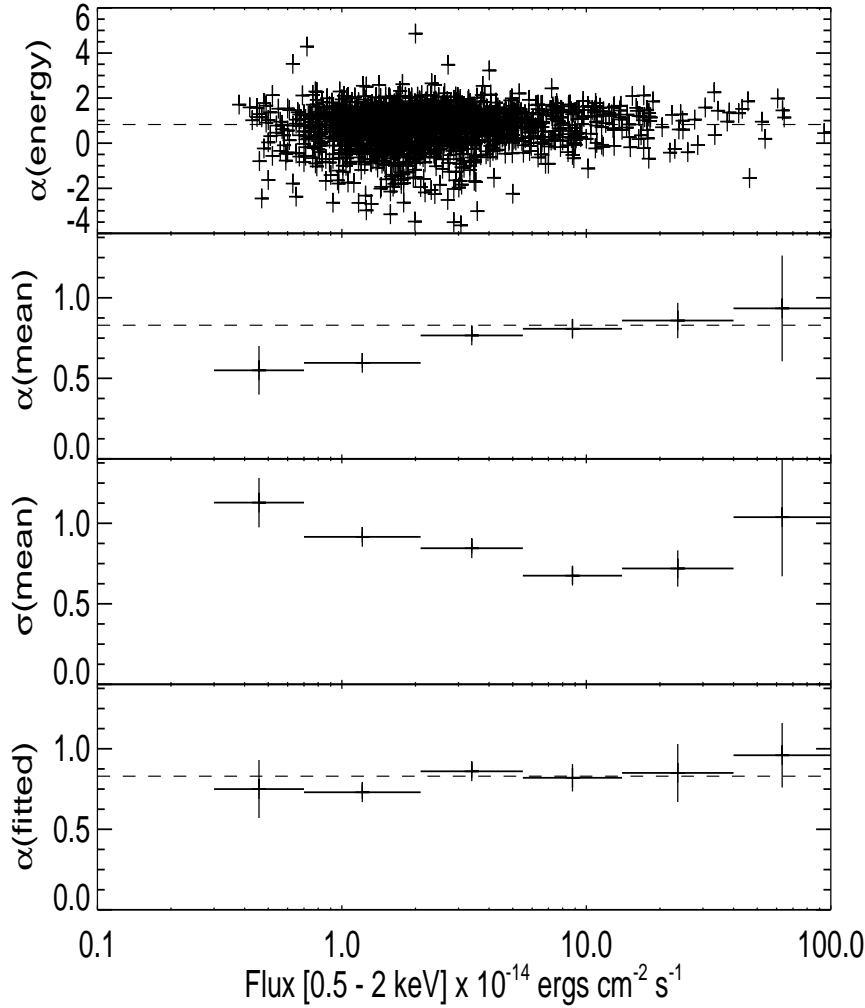,height=15truecm,width=15truecm,angle=0}

\caption{Fits to the whole RIXOS dataset, consisting of 1762 objects.  The
top panel shows the power-law slope fit to all the data assuming galactic
$N_H$.  The second panel shows the data averaged in 6 flux bins, together with
the standard error.  A clear trend can be seen, with the average hardening with
lower fluxes.  The third panel shows the standard deviation of the data around
the mean.  The bottom panel shows the fitted average slope and error in each
flux bin.  This method essentially biases against extreme outliers, and is more
representative of the mode of the distribution.  These data show no strong
correlation with flux. In all cases the dashed line is representative of the
mode of the slopes.}

\label{flux_slope}
\end{figure*}

Finally, we have analysed the whole RIXOS sample, including all
sources in all RIXOS fields down to the detection limit of each field,
containing 1762 sources.  Even though we do not have identifications
for most of the sources with a flux below $3 \times 10^{-14}$ ergs
cm$^{-2}$ s$^{-1}$ we can still study the spectral shape of the
faintest sources which are precisely those which will contribute most
to the soft X-ray background.  Recent work has indicated that there
may be a correlation between the average spectral slope and flux
(Hasinger et al. 1993, Vikhlinin et al. 1995).  Vikhlinin et
al. (1995) analysed 130 ROSAT fields and extracted average spectra
over a range of flux bins. They showed a correlation between source
flux and spectral index, with bright sources ($> 2 \times 10^{-13}$
ergs cm$^{-2}$ s$^{-1}$) having average slopes close to 1.3, and faint
sources ($< 10^{-14}$ ergs cm$^{-2}$ s$^{-1}$) having average slopes
close to 0.5. As noted by Vikhlinin et al., the average slope of 0.5
is close to that obtained for the soft X-ray background.


The exact significance of this correlation is, however, unclear since Vikhlinin
et al. used either hardness ratios, which in the case of the faintest sources
will have a bias to softer slopes (see section \ref{cash_test}), or summed up
all the sources in a flux bin and fitted multi-channel data with  models using
$\chi^2$.  While summing up the data will allow higher signal-to-noise at a
higher resolution, it can only give information on the average properties of
the sources and not on the distribution of slopes in a given flux band.  

By using the Cash statistic technique on three-colour data we can avoid
problems of biases.  We have analysed the whole RIXOS sample containing 1762
sources including sources which extend down to a flux of $4 \times 10^{-15}$
ergs cm$^{-2}$ s$^{-1}$. Figure~\ref{flux_slope} shows the fits to the whole
sample as a function of flux both as a scatter plot and binned into flux bins.
The second panel shows the average of the slopes in each flux bin which is the
equivalent of the Vikhlinin et al. data.  As with the sample of Vihklinin et
al. there is a clear trend towards harder spectra at lower flux limits in the
latter.  Above the RIXOS flux limit there is no significant hardening, while
there is a significant deviation below $\sim 2 \times 10^{14}$ ergs cm$^{-2}$
s$^{-1}$.  However, unlike the Vihklinin sample we can look at the distribution
of slopes within a flux bin. The top panel of Figure~\ref{flux_slope} makes it
clear that the majority of sources do not show a trend to harder slopes at
lower flux limits. To highlight this further, we have fitted the mean slope and
dispersion within each flux bin in the same way we have done for the RIXOS
sample (section~6).  The bottom panel shows the fitted average slope as a
function of flux and it is clear that there is no significant trend.  The most
obvious explanation for this discrepancy between the arithmetically averaged
data and the fitted average value is that there are changes in the distribution
of slopes within a flux bin rather than a global change in the spectral slope.
This is supported by the third panel which shows an increase in the measured
standard deviation of the data with decreasing flux.  In the RIXOS data it is
clear that there are a number of very hard sources below a flux of $3 \times
10^{-14}$ ergs cm$^{-2}$ s$^{-1}$.  These sources would bias the mean but, as
outliers, they would not have a significant effect on fits of a gaussian
distribution to the slopes, exactly as observed.  If we look at those sources
where $\alpha < 0$, then 20\% of sources below $3 \times 10^{-14}$ ergs
cm$^{-2}$ s$^{-1}$ satisfy this condition.  However, if we look at all those
sources with fluxes greater than $3 \times 10^{-14}$ ergs cm$^{-2}$ s$^{-1}$,
then only 13\% of the sources satisfy this criterion.  It is therefore clear
that as we go to fainter fluxes a higher proportion of the sources are very
hard.  From studies of deep ROSAT pointings it is unlikely that there are a
significant number of stars at faint fluxes, and it is more likely that the
sources with hard spectra are an absorbed population. One obvious candidate for
these sources are Seyfert 2 galaxies.  Such sources are both hard and faint
relative to un-obscured AGN and would therefore give the observed distribution
which has more hard sources at fainter fluxes.

\subsection{The nature of the soft X-ray background}

One question that has been highlighted by recent ROSAT observations is the
nature of the soft X-ray background.  It has been known for many years that the
average slope of the soft X-ray background is $\alpha=0.4$ (for a review see
Fabian and Barcons 1992).  This spectrum is inconsistent with the average
spectrum of (relatively bright) AGN.  From the deepest surveys undertaken to
date it is clear that a significant fraction of the X-ray background is made up
of emission from narrow emission line galaxies (NELGs), an amorphous
classification that may include Seyfert 2 galaxies as well as starburst
galaxies and LINERS.   However, the average spectrum of the RIXOS NELGs is also
too soft to explain the soft X-ray background. 

In contrast, the mean spectrum of NELGs at fainter flux levels {\it is}
consistent with the X-ray background (Romero-Colmenero et al. 1996) and harder
than that of AGN even in the same flux range. In the previous section, we
showed that the hard overall mean spectrum  at faint fluxes is caused by a
population of very hard sources (i.e. harder than the background). These  bias
the mean source spectral slope to a value which is consistent with the slope of
the background. 

A combination of two possible explanations 
may account for the slope of the X-ray background, given that the NELGs
are likely to be a mixture of intrinsic source types. There may be a genuine
change in the dominant emission mechanism between the bright RIXOS NELGs and
those identified in deep surveys, with the emission from faint NELGs being
dominated by a hot continuum source (for example hot gas, perhaps associated
with an extended halo rather than the galactic nucleus); or the very hard
sources which we identify in the extended RIXOS sample may be an absorbed
population consisting, say, of Seyfert 2 galaxies (cf. Grindlay and Luke
1990).  More sensitive individual X-ray  spectral observations of a sample of
faint sources will be needed to resolve this question.

\section{Conclusions}

RIXOS is a flux limited, nearly complete sample of X-ray selected sources.  We
have demonstrated that for such a sample  it is possible to obtain useful
spectral information even down to very faint limits, as long as the correct
statistic is used. In contrast, a simple  hardness ratio method, which has been
used by a number of authors to determine the spectral slope for faint sources,
is shown to bias in the inferred power-law slope towards a
steeper spectrum. The use of three-colour data allows some discrimination
between thermal and non-thermal X-ray emission, at least for
relatively bright sources.  We have determined the spectral characteristics for
each sub-category of sources within the RIXOS survey:

\begin{itemize}

\item[1)] Though little can be said directly about the X-ray spectra of the
stars, the use of three-colour data demonstrates the ability of the method to
discriminate between thermal and non-thermal sources.

\item[2)] Most of the RIXOS clusters are consistent with the majority of the
emission arising from hot ($> 3$ keV) gas.  There are some clusters where there
is evidence for a lower temperature, which may indicate the presence of a
cooling flow. 

\item[3)] On average, the NELGs have X-ray spectra that are consistent with the
spectra of the AGN and may indicate that many of the NELGs found in RIXOS are,
in fact, low-luminosity AGN.  This is at variance with the X-ray spectra of
NELGs found from deep X-ray surveys, where the average slope is much harder.
The NELGs observed in deep X-ray surveys are then either a more absorbed
population of sources, or the X-ray emission in the faintest NELGs arises from
some other mechanism other than an AGN non-thermal power-law.

\item[4)] The AGN have an average slope of $\alpha = 1.05 \pm 0.05$ with no
evidence for spectral evolution.  This average slope is somewhat harder than
the averages found for other samples of soft X-ray selected AGN.  However, the
inappropriate use of hardness ratios will have softened the average slopes in
other samples.  The value of $\alpha = 1.05$ is consistent with the naked
power-law expected from AGN, implying that the X-ray spectrum of the RIXOS AGN
is relatively uncontaminated by processes such as reflection and absorption.
Since many of the previous X-ray selected samples concentrated on low
redshift/low luminosity AGN, part of the discrepancy between RIXOS and other
samples may be ascribed to the effect of redshift and/or luminosity on the
processes that modify the underlying power-law.

\item[5)] Analysis of the whole RIXOS sample confirms the presence of a flux
dependent spectral slope (Hasinger et all 1993, Vikhlinin et al. 1995).
However, we have been able to investigate the cause of this correlation and the
most likely explanation is of an increasing proportion of very hard sources
rather than an average hardening of the spectra.

\end{itemize}

\section*{Acknowledgments}

We thank all the RIXOS team for their work in obtaining and reducing
the data.  We would also like to thank Liz Puchnarewicz for checking
the source count extraction procedure.  FJC thanks the DGES for
parcial financial support, under project PB95-0122. H.A. benefited
from financial support by CONACYT (Mexico; C\'atedra Patrimonial, ref
950093).  The RIXOS project has been made possible by the award of
International Time on the La Palma telescopes by the Comit\'e
Cient\'\i fico Internacional.  This research has made use of data
obtained from the UK {\it ROSAT} Data Archive Centre at the Department
of Physics and Astronomy, University of Leicester (LEDAS).  We thank
the Royal Society for a grant to purchase equipment essential to the
RIXOS project.

\begin{table*}
\vbox to220mm{\vfil Landscape table to go here.
\vfil}
\end{table*}

\begin{table*}
\vbox to220mm{\vfil Landscape table to go here.
\vfil}
\end{table*}

\begin{table*}
\vbox to220mm{\vfil Landscape table to go here.
\vfil}
\end{table*}

\begin{table*}
\vbox to220mm{\vfil Landscape table to go here.
\vfil}
\end{table*}

\begin{table*}
\vbox to220mm{\vfil Landscape table to go here.
\vfil}
\end{table*}

\begin{table*}
\vbox to220mm{\vfil Landscape table to go here.
\vfil}
\end{table*}

\begin{table*}
\vbox to220mm{\vfil Landscape table to go here.
\vfil}
\end{table*}

\begin{table*}
\vbox to220mm{\vfil Landscape table to go here.
\vfil}
\end{table*}

\begin{table*}
\vbox to220mm{\vfil Landscape table to go here.
\vfil}
\end{table*}

\begin{table*}
\vbox to220mm{\vfil Landscape table to go here.
\vfil}
\end{table*}

\begin{table*}
\vbox to220mm{\vfil Landscape table to go here.
\vfil}
\end{table*}

\begin{table*}
\vbox to220mm{\vfil Landscape table to go here.
\vfil}
\end{table*}

\begin{table*}
\vbox to220mm{\vfil Landscape table to go here.
\vfil}
\end{table*}


\begin{thebibliography}{}
\bibitem[\protect\citename{Almaini et al.}1996]{almaini}
Almaini O., Shanks T., Boyle B.J., Griffiths R.E., Roche N., Stewart G.C., 
Georgantopoulos I., 1996, MNRAS, 282, 295
\bibitem[\protect\citename{Bade et al.}1996]{bade}
Bade N., Komossa, S., Dahlem, M., 1996, A\&A, 309, 35L
\bibitem[\protect\citename{Bechtold et al.}1994]{bechtold}
Bechtold, J., Elvis, M., Fiore, F., Kuhn, O., Cutri, R.M., McDowell, J.C., 
Rieke, M., Siemiginowska, A., Wilkes, B.J., 1994, AJ, 108, 759 
\bibitem[\protect\citename{Boller et al.}1994]{boller}
Boller, T., Fink, H., Schaeidt, S., 1994, A\&A, 291, 403
\bibitem[\protect\citename{Bowyer et al.}1994]{bowyer}
Bowyer S., Lieu R., Lampton M., Lewis J., Wu X., Drake J.J., Malina R., 1994,
ApJS, 93, 569
\bibitem[\protect\citename{Boyle et al.}1994]{boyle}
Boyle, B.J., Shanks, T., Georgantopoulos, I., Stuart, G.C., Griffiths, R.E.,
1994, MNRAS, 271, 639
\bibitem[\protect\citename{Boyle et al.}1995]{boyle2}
Boyle, B.J., McMahon, R.G., Wilkes, B.J, Elvis, M.,
1995, MNRAS, 272, 462
\bibitem[\protect\citename{Branduardi-Raymont et al.}1995]{branduardi-ray}
Branduardi-Raymont G., Mason K.O., Warwick R.S., Carrera F.J., Graffagnino
V.G., Mittaz J.P.D.,E Puchnarewicz E.M., Smith P.J., Barber C.R., Pounds K.A.,
Stewart G.C., McHardy I.M., Jones L.R., Merrifield M.R., Fabian A.C., McMahon
R.G., Ward M.J., George I.M., Jones M.H., Lawrence A., Rowan-Robinson M., 1994,
MNRAS, 270, 947
\bibitem[\protect\citename{Carrera}1998]{carrera}
Carrera F.J., 1998, MNRAS, submitted
\bibitem[\protect\citename{Cash}1979]{cash}
Cash J., 1979, ApJ, 228, 939
\bibitem[\protect\citename{Castander et al.}1995]{castander}
Castander F.J., Bower R.G., Ellis R.S., Aragon-Salamanca A., Mason K.O.,
Hasinger G., McMahon R.G., Carrera F.J., Mittaz J.P.D., P\'erez--Fournon I.,
Lehto H,J., 1995, Nature, 377, 39
\bibitem[\protect\citename{Ciliegi et al.}1996]{ciliegi}
Ciliegi P., Elvis, M., Wilkes, B.J., Boyle, B.J., McMahon, R.G., 1996, 
MNRAS, 284, 401 
\bibitem[\protect\citename{Elvis et al.}1986]{elvis}
Elvis, M., Green, R.F., Bechtold, J., Scmidt, M., Neugebauer, G., Soifer, B.T.,
Mathews, K., Fabbiano, G., 1986, ApJ, 310, 291
\bibitem[\protect\citename{Hasinger et al.}1993]{hasinger}
Hasinger G, Burg R., Giacconi R., Hartner G., Schmidt M., Trumper J.,
 Zamorani G., 1993, A\&A, 275, 1
\bibitem[\protect\citename{Fabian and Barcons}1992]{fabian}
Fabian, A.C. and Barcons, X., 1992, ARA\&A, 30, 429
\bibitem[\protect\citename{Fiore et al.}1993]{fiore93}
Fiore, F., Elvis, M., Mathur, M., Wilkes, B.J., McDowell, J.C., 1993, 415, 129
\bibitem[\protect\citename{Fiore et al.}1994]{fiore}
Fiore, F., Elvis, M., McDowell, J.C., Siemiginowska, A., Wilkes, B.J., 1994,
ApJ, 431, 515
\bibitem[\protect\citename{Gioia et al.}1990]{gioia}
Gioia I. M., Maccacaro T., Schild R. E., Wolter A., Stocke J. T., Morris
 S. L., Henry J. P., 1990, ApJS, 72, 567
\bibitem[\protect\citename{Giommi et al.}1991]{giommi}
Giommi P., Tagliaferri G., Beuermann K., Branduardi-Raymont G., Brissenden
 R., Graser U., Mason K.  O., Mittaz J. P. D., Murdin P., Pooley G.,
 Thomas H. -C., Tuohy I., 1991, ApJ, 378, 77
\bibitem[\protect\citename{Grindlay and Luke} 1990]{grindlay}
Grindlay, J.E. and Luke, M., 1990, in: High resolution X-ray spectroscopy of
cosmic plasmas, Cambridge University Press, 276
\bibitem[\protect\citename{Koratkar et al.} 1995]{koratkar}
Koratkar A., Deustua S.E., Heckman T., Filippenko A.V., Ho L.C., Rao M., 1995,
ApJ, 440, 132
\bibitem[\protect\citename{Loar et al.}1994]{laor}
Laor, A., Fiore, F., Elvis, M., Wilkes, B.J., McDowell, J.C., 1994, ApJ, 435,
611 
\bibitem[\protect\citename{Loredo}1990]{loredo}
Loredo T.J., 1990, in: Maximum Entropy and Bayesian Methods, Ed. P.F. Foug\'ere, 81
\bibitem[\protect\citename{Maccacaro et al.}1988]{maccacaro}
Maccacaro T., Gioia I.M., Wolter A., Zamorani G., Stocke J.T., 1988, ApJ, 326,
680
\bibitem[\protect\citename{Mason et al.}1997]{m96}
Mason K.O. et al., 1998, in preparation
\bibitem[\protect\citename{McHardy et al.}1998]{mchardy}
McHardy I.M. et al., 1997, MNRAS, in press.
\bibitem[\protect\citename{Mittaz et al.}1997]{mittaz}
Mittaz, J.P.D. et al., 1998, in preparation
\bibitem[\protect\citename{Nandra \& Pounds}1992]{np92}
Nandra, K. and Pounds, K.A., 1992, Nature, 359, 215
\bibitem[\protect\citename{Nandra \& Pounds}1994]{nandra}
Nandra K. \& Pounds, K.A., 1994, MNRAS, 268, 405
\bibitem[\protect\citename{Pan, Stewart \& Pounds}1990]{pan}
Pan, H.C., Stewart, G.C., Pounds, K.A., 1990, MNRAS, 242, 177
\bibitem[\protect\citename{Page et al.}1996]{page}
Page M.J., Carrera, F.J., Hasinger, G., Mason, K.O., McMahon, R., Mittaz,
J.P.D., Barcons, X., Carballo, R., Gonzalez-Serrano, I., P\'erez--Fournon, I., 
1996, MNRAS, 281, 579
\bibitem[\protect\citename{Pounds et al.}1993]{pounds}
Pounds K.A. et al., 1993, MNRAS, 260, 77 
\bibitem[\protect\citename{Press et al.}1992]{press}
Press W.H., Teukolsky S.A., Vetterling W.T., Flannery B.P., 1992, 
Numerical Recipes, Cambridge University Press, 617
\bibitem[\protect\citename{Puchnarewicz et al.}1996a]{p96a}
Puchnarewicz E.M., Mason K.O., Romero-Colmenero E., Carrera F.J., 
Hasinger G., McMahon R., Mittaz J.P.D., Page M.J., Carballo R., 1996,
MNRAS, 281, 1243
\bibitem[\protect\citename{Puchnarewicz et al.}1996]{p97}
Puchnarewicz, E. M., Mason, K. O., Carrera, F. J., Brandt, W. N.,
Cabrera-Guera, F., Carballo, R., Hasinger, G., McMahon, R. G.,
Mittaz, J. P. D., Page, M. J., Perez-Fournon, I., Schwope, A., 1997,
MNRAS, 291, 177
\bibitem[\protect\citename{Reimers et al.}1995]{reimers}
Reimers, D., Bade, N., Schartel, N., Hagen, H.-J., Engels, D, Toussaint, F.,
1995, A\&A, 296, 49
\bibitem[\protect\citename{Romero-Colmenero et al.}1996]{rom2}
Romero-Colmenero E. Branduardi-Raymont, G., Carrera, F.J., Jones, L.R., Mason,
K.O., McHardy, I.M., Mittaz, J.P.D., 1996, MNRAS, 282, 94
\bibitem[\protect\citename{Romero-Colmenero et al.}1997]{rom3}
Romero-Colmenero E. et al., 1997, in preparation
\bibitem[\protect\citename{Sarazin}1986]{sarazin}
Sarazin C.L., 1986, Reviews of Modern Physics, 58, 1
\bibitem[\protect\citename{Schmitt et al.}1990]{schmitt}
Schmitt J.H.M.M., Collura A., Sciortino S., Vaiana G.S., Harnden F.R., 
Rosner R., 1990, ApJ, 365, 704
\bibitem[\protect\citename{Schartel et al.}1996]{schartel}
Schartel, N., Walter, R., Fink, H,H., Tr\"umper, J., 1996, A\&A, 307, 33
\bibitem[\protect\citename{Snowden et al.}1994]{snowden}
Snowden S.L., McCammon D., Burrows D.N., Mendenhall J.A., 1994, ApJ, 424, 714
\bibitem[\protect\citename{Thompson \& Cordova}1994]{thompson}
Thompson, R.J. and Cordova, F.A., 1994, ApJ, 434, 54
\bibitem[\protect\citename{Turner \& Pounds}1988]{turner}
Turner, T.J. and Pounds, K.A., 1988, MNRAS, 232, 463
\bibitem[\protect\citename{Vihklinin et al.}1995]{vikhlinin}
Vikhlinin, A., Forman, W., Jones, C., Murray, S., 1995, ApJ, 451, 564
\bibitem[\protect\citename{Walter \& Fink}1993]{wf}
Walter, R. \& Fink, H.H., 1993, A\&A, 274, 
\end{thebibliography}
\end{document}